\documentclass[apjl]{emulateapj}

\newcommand{\ctbd}[1]{}

\newcommand{\cfa}{Harvard-Smithsonian Center for Astrophysics (CfA)}

\newcommand{\ghr}{\ensuremath{^h}}
\newcommand{\gmin}{\ensuremath{^m}}

\newcommand{\kms}{\ensuremath{\rm km\,s^{-1}}}
\newcommand{\ms}{\ensuremath{\rm m\,s^{-1}}}

\newcommand{\gcmc}{\ensuremath{\rm g\,cm^{-3}}}

\newcommand{\teff}{\ensuremath{T_{\rm eff}}}
\newcommand{\logg}{\ensuremath{\log{g}}}
\newcommand{\vsini}{\ensuremath{v \sin{i}}}
\newcommand{\feh}{[Fe/H]}

\newcommand{\rsun}{\ensuremath{R_\sun}}
\newcommand{\msun}{\ensuremath{M_\sun}}
\newcommand{\lsun}{\ensuremath{L_\sun}}

\newcommand{\rstar}{\ensuremath{R_{\star}}}

\newcommand{\lstar}{\ensuremath{L_{\star}}}

\newcommand{\rpl}{\ensuremath{R_{p}}}
\newcommand{\mpl}{\ensuremath{M_{p}}}

\newcommand{\rhopl}{\ensuremath{\rho_{p}}}
\newcommand{\ipl}{\ensuremath{i_{p}}}

\newcommand{\rjup}{\ensuremath{R_{\rm J}}}
\newcommand{\mjup}{\ensuremath{M_{\rm J}}}

\newcommand{\figr}[1]{Fig.~\ref{fig:#1}}
\newcommand{\secr}[1]{\mbox{\S\ \ref{sec:#1}}}

\newcommand{\tabr}[1]{\mbox{Table~\ref{tab:#1}}}

\newcommand{\flwof}{\mbox{FLWO 1.2 m}}

\newcommand{\flwos}{\mbox{FLWO 1.5 m}}

\shorttitle{HAT-P-4b: A low-density transiting planet}
\shortauthors{Kov\'acs et al.}

\begin{document}

\title{HAT-P-4\lowercase{b}: A metal-rich low-density transiting hot Jupiter\altaffilmark{\dag}}
\author{
	G.~Kov\'acs\altaffilmark{1},
	G.~\'A.~Bakos\altaffilmark{2,3},
	G.~Torres\altaffilmark{2},
	A.~Sozzetti\altaffilmark{2,4},
	D.~W.~Latham\altaffilmark{2},
	R.~W.~Noyes\altaffilmark{2},
	R.~P.~Butler\altaffilmark{5},
	G.~W.~Marcy\altaffilmark{6},
	D.~A.~Fischer\altaffilmark{7},
	J.~M.~Fern\'andez\altaffilmark{2},
	G.~Esquerdo\altaffilmark{2},
	D.~D.~Sasselov\altaffilmark{2},
	R.~P.~Stefanik\altaffilmark{2},
	A.~P\'al\altaffilmark{8},
        	J.~L\'az\'ar\altaffilmark{9},
        I.~Papp\altaffilmark{9}, \&
        P.~S\'ari\altaffilmark{9}
}

\altaffiltext{1}{Konkoly Observatory, Budapest, P.O.~Box 67, H-1125, Hungary; 
                 kovacs@konkoly.hu}
\altaffiltext{2}{\cfa, 60 Garden Street, Cambridge, MA 02138, USA}
\altaffiltext{3}{Hubble Fellow.}
\altaffiltext{4}{INAF - Osservatorio Astronomico di Torino, 
	Strada Osservatorio 20, 10025, Pino Torinese, Italy}
\altaffiltext{5}{Department of Terrestrial Magnetism, Carnegie  
	Institute of Washington DC, 5241 Broad Branch Rd.~NW, Washington
	DC, USA 20015-1305}
\altaffiltext{6}{Department of Astronomy, University of California,
	Berkeley, CA 94720, USA}
\altaffiltext{7}{Department of Physics \& Astronomy, San Francisco
	State University, San Francisco, CA 94132, USA}
\altaffiltext{8}{Department of Astronomy, E\"otv\"os Lor\'and University, 
        Budapest, Hungary}
\altaffiltext{9}{Hungarian Astronomical Association, Budapest, Hungary}
\altaffiltext{\dag}{
	Based in part on observations obtained at the W.~M.~Keck
	Observatory, which is operated by the University of California and
	the California Institute of Technology. Keck time has been in part
	granted by NASA.
}


%
%
\begin{abstract}
We describe the discovery of HAT-P-4b, a low-density extrasolar 
planet transiting BD+36~2593, a $V=11.2$\,mag slightly evolved 
metal-rich late F star. The planet's orbital period is 
$3.056536\pm0.000057$\,d with a mid-transit epoch of 
$2,454,245.8154\pm0.0003$ (HJD). Based on high-precision 
photometric and spectroscopic data, and by using transit light 
curve modeling, spectrum analysis and evolutionary models, we 
derive the following planet parameters:
\mpl$=0.68\pm0.04$\,\mjup, 
\rpl$=1.27\pm0.05$\,\rjup,
\rhopl$=0.41\pm0.06$\,\gcmc\ and 
$a=0.0446\pm0.0012$\,AU\@.
Because of its relatively large radius, together with its assumed 
high metallicity of that of its parent star, this planet adds to 
the theoretical challenges to explain inflated extrasolar planets. 
\end{abstract}


%
%
\section{Introduction}
\label{sec:intro}
%
%
In the course of our ongoing wide field planetary transit search 
program HATNet \citep{bakos04}, we have discovered a large radius 
and low density planet orbiting an $11$th magnitude star BD+36~2593. 
This planet is the fifth member of a group of low-density transiting 
exoplanets. The combination of its low mass and the relatively 
high metallicity and age of the parent star makes theoretical 
interpretation of its large radius difficult. In this Letter we 
describe the observational properties of the system and derive the 
physical parameters both for the host star and for the planet. We also 
briefly comment on the theoretical status of inflated extrasolar planets.

%
%
\section{The photometric discovery and follow-up observations}
\label{sec:phot}

The star BD+36~2593 (also GSC~02569-01599) at 
$\alpha = 15\ghr19\gmin57\fs92$, $\delta = +36\degr13\arcmin46\farcs7$, 
is contained in field G191 of HATNet, centered at 
$\alpha = 15\ghr28\gmin$, $\delta = +37\degr30\arcmin$.
As we show in the remainder of the paper, the star is orbited by a
planetary companion, and so we label the host star as HAT-P-4 and
the planet as HAT-P-4b. Field G191 was monitored from 2004 December 
until 2005 June by the HAT-7 telescope at the Fred Lawrence Whipple 
Observatory (FLWO) of the Smithsonian Astrophysical Observatory (SAO). 
Nearly $90$\% of the data points were gathered after February 2005. 
The field is relatively sparse, with $\sim 14000$ objects brighter than 
$I\approx14$\,mag. Most of the light curves have some $5300$ data points. 
The sampling cadence is $5.5$\,min, slightly longer than the $5$\,min 
integration time.

\begin{figure}
\epsscale{1.0}
\plotone{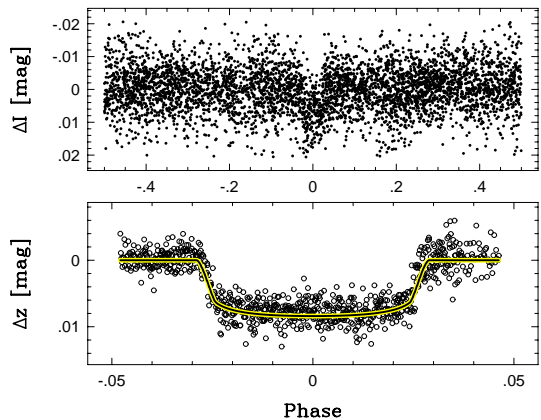}
\caption{
	Folded and unbinned light curves of HAT-P-4. The HATNet and 
	\flwof\ data are plotted in the upper and lower panels, 
	respectively. The transit model fit to the FLWO data is shown 
	by continuous line (see \tabr{planet} for the resulting transit 
	parameters). The out-of-transit level of the HATNet data 
	corresponds to $I=10.537$\,mag, phase is zero at the transit 
	center.} 
\label{fig:lcs}
\end{figure}

%
\notetoeditor{This is the intended place of \figr{lcs}}

A first look at the compact subset collected in the $98$\,d time 
span between March and June, 2005, revealed the presence of a transit
signal with a nearly integer $3$\,d period. The detection was made  
possible by the application of the Trend Filtering Algorithm 
\citep[TFA;][]{kovacs05}. However, assuming the primary is a 
main-sequence F or G type star, the relative length of the transit 
is some $20-30$\% greater than expected for a Jupiter-size companion 
with the above orbital period. Although both the lack of detection 
in the raw data and the length of the transit made us suspicious about 
the viability of this candidate, we left some room for the possibility 
that the primary was slightly evolved (an assumption that proved to 
be true in the subsequent investigations --- see \secr{params}). 
Therefore, we proceeded with rejection-mode spectroscopy, which showed 
no sign of radial velocity (RV) variation at the \kms\ level. 

An improved reduction of all available frames of the field was
completed by February 2007. In addition to using a new data pipeline
that employs refined astrometry \citep{pal06} and aperture photometry,
we detrended the light curves before signal search with the aid of an
External Parameter Decorrelation technique (EPD, see also Bakos et
al.~2007). This technique utilizes the fact that various ``external
parameters'' that are specific to the star, such as sub-pixel position
on the frame, point spread function properties (e.g., width and
elongation), or specific to the frame, such as telescope position, 
are correlated with the deviations of the star's brightness from the
median. The technique is optimal for stars that are otherwise
non-variable most of the time (e.g.~transit candidates). EPD derives 
the correlation between brightness deviations and the underlying 
external parameters, and subsequently corrects them for each individual 
star. This is different from what is done during the application of 
TFA, where we consider the full time history of the light variation 
and use the light curves to ``cure themselves'' by recognizing the 
hidden systematics in each other. The two methods are complementary 
and we use both (EPD followed by TFA). All these led to a powerful 
confirmation of the earlier detection.

The folded light curve constructed from the HATNet data is shown in the
upper panel of \figr{lcs}. The best period was obtained from the BLS
analysis \citep[][]{kovacs02} after applying TFA with some 700 template 
stars.\footnote{The EPD technique played the main role in the detection; 
TFA subsequently led to an increase of $20$\% in the signal-to-noise 
ratio.} As seen, the transit is clearly visible in the light curve even 
though no binning was applied. There is no sign of periodic out-of-transit 
variation, indicating that this is a near ``textbook'' transit signal. 

To update the photometric ephemeris and obtain a precise light curve
for model fitting, we observed our target with KeplerCam on the
$1.2$\,m telescope at FLWO\@. We observed two full transits on May 
24 and 27, 2007. The folded light curve in the Sloan $z$ band for 
all $985$ data points is shown in the lower panel of \figr{lcs}. 
The follow-up observations verify the discovery light curve and 
are accurate enough to make transit modeling possible. 

%
%
\vspace{-2mm}
\section{The spectroscopic verification and exclusion of blend 
scenarios}
\label{sec:spectr}

As mentioned in \secr{phot}, basic rejection mode spectroscopy to 
exclude stellar binary and certain blend types was already initiated 
after the first detection in the fall of 2005. By June 2006 we had 
collected 9 spectra with the CfA Digital Speedometer 
\citep[DS;][]{latham92} on the \flwos\ telescope, covering a 45\,\AA\ 
window centered at the Mg~b triplet at 5187\,\AA\@. We derived spectroscopic 
parameters by comparing the observed spectra against synthetic spectra, 
as described in detail by \citet{torres02}. The initial stellar 
parameters from this analysis were: \teff$=5500$\,K, \logg$=3.5$\,[cgs] and 
\vsini$=5.9$\,\kms, indicating an evolved G-type primary\footnote{These 
parameters were derived with [Fe/H]$=0.0$. By using [Fe/H]$=0.24$ from 
the Keck spectra we get a slightly better match with the synthetic 
spectra that yields \teff$=6000$\,K and \logg$=4.0$, in better 
agreement with the Keck results described below.}. The DS spectra were 
also used to derive radial velocities by cross-correlating the spectra 
with synthetic spectra based on Kurucz model atmospheres 
\citep[e.g.][]{latham02}. The rms scatter of the RV data was $0.71$\,\kms 
around the mean velocity of $-1.58$\,\kms. The velocities did not show 
any correlation with the orbital phase and we found no sign of a second 
stellar component in the spectra. These spectroscopic results, together 
with the secure detection in the HATNet data, made a viable case for 
obtaining high-precision RV measurements to seek evidence for orbital 
motion and to obtain refined stellar parameters. Accordingly, 
high-resolution spectroscopy was conducted with the HIRES instrument 
\citep{vogt94} on the Keck~I telescope between 2007 March 27 and May 29. 
See, e.g., \cite{torres07} for the details of the observational procedure. 
The nine resulting RV measurements are listed in \tabr{rv}. They are 
relative velocities in the Solar System barycentric frame of reference 
\citep[see][]{butler96}.

%
%
\notetoeditor{This is the intended place of \tabr{rv}.}

\begin{deluxetable}{crcrc}
\tabletypesize{\scriptsize}
\tablewidth{0pt}
\tablecaption{
	\label{tab:rv}
	HIRES Relative Radial Velocities for HAT-P-4.}
\tablewidth{0pt}
\tablehead{
	\colhead{BJD} &
	\colhead{RV}  &
	\colhead{$\sigma_{\rm{RV}}$} &
	\colhead{O-C} &
	\colhead{Phase}\tablenotemark{a}\\
	\colhead{($2,\!400,\!000+$)} &
	\colhead{(\ms)} &
	\colhead{(\ms)} & 
	\colhead{(\ms)} & 
	\colhead{}
	}
\startdata
    54186.98522 & $ 101.0\ \ \ $ & 2.1 & $  2.7\ \ \ $ & 0.753\\
    54187.11241 & $  92.6\ \ \ $ & 2.1 & $ -2.6\ \ \ $ & 0.794\\
    54188.01160 & $ -25.4\ \ \ $ & 2.0 & $  0.3\ \ \ $ & 0.088\\
    54188.07150 & $ -35.8\ \ \ $ & 2.0 & $ -2.0\ \ \ $ & 0.108\\
    54189.00174 & $ -28.1\ \ \ $ & 2.2 & $ -2.8\ \ \ $ & 0.412\\
    54189.08262 & $ -13.9\ \ \ $ & 2.0 & $ -0.7\ \ \ $ & 0.439\\
    54189.13222 & $  -3.1\ \ \ $ & 3.3 & $  2.3\ \ \ $ & 0.455\\
    54249.93769 & $ -52.5\ \ \ $ & 2.8 & $ -3.6\ \ \ $ & 0.349\\
    54279.86357 & $ -36.8\ \ \ $ & 2.9 & $  8.4\ \ \ $ & 0.139\\
\enddata
\tablenotetext{a}{Relative to the center of transit.}

\end{deluxetable}

\begin{figure}
\epsscale{1.0}
\plotone{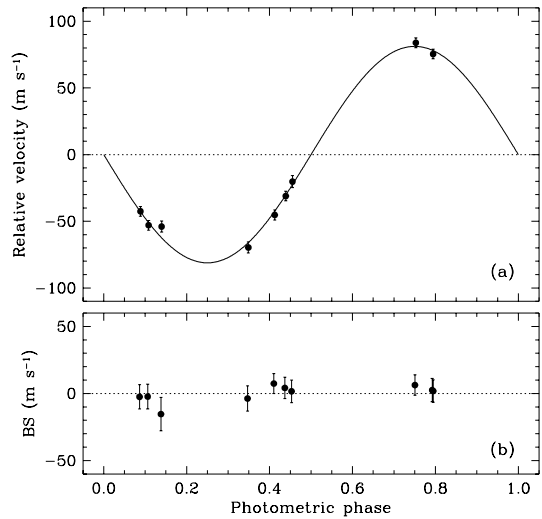}
\caption{
	Relative radial velocity and bisector span variations from 
	the Keck observations of HAT-P-4. Panel (a) shows the nine Keck 
	RV measurements with a zero-eccentricity orbital fit. In panel 
	(b) we show the bisector spans (from nine iodine exposures and 
	one template spectrum) displaying a much smaller scatter. Error 
	bars include the 3~\ms\ estimated velocity jitter.}	
\label{fig:rv}
\end{figure}

%
\notetoeditor{This is the intended place of \figr{rv}}

The data were fitted by a circular\footnote{A fit with free-floating 
eccentricity yields $e=0.055\pm0.035$, supporting the assumption of a 
circular orbit.} orbit with period and mid transit epoch constrained 
by the photometric data (see \tabr{planet}). The result of this 
two-parameter (systematic offset and semi-amplitude) fit is displayed 
in \figr{rv}. The unbiased estimate of the standard deviation of the 
residuals is $4.1$\,\ms, larger than the internal errors listed in 
\tabr{rv}. A part of this scatter may come from the last data point 
that was taken under unfavorable weather conditions (leaving out this 
point we get $2.7$\,\ms\ for the standard deviation of the residuals). 
Nevertheless, it is also possible that there is a ``velocity jitter'' 
due to stellar activity. The size of this jitter is estimated to be 
about $3$\,\ms, falling in the expected range for an inactive late F-type 
star \citep[][]{wright05}. The low activity level is supported by 
the absence of emission features in the Ca~II H and K lines. 

A crucial part of the verification of the sub-stellar nature of the
companion is testing various blend scenarios. There are at least two 
basic ways of such testing: (i) light curve modeling, combined with 
the information available from the spectra (e.g., maximum flux 
contribution by the blended binary); (ii) checking spectral line 
asymmetry by searching for bisector span variations 
\citep[e.g.,][]{santos02,torres05}. Our experience shows that 
while (i) requires several pieces of information (good quality full 
light curve, running many binary models and performing a fairly deep 
spectrum analysis), (ii) is more sensitive to hidden components and 
simpler to perform. Therefore, we settled on (ii) and searched for 
a variation in the differences measured between the velocities at the 
top and the bottom of the correlation profiles of the individual spectra. 
The result is shown in the lower panel of \figr{rv}. For a hidden 
stellar binary blended by our target we would expect a variation in the
bisector span in phase with the radial velocity, and with an amplitude
comparable to that of the measured RV variation itself \citep{torres05}. 
However, the standard deviation of the bisector variation is $6.5$\,\ms,
which is only $8$\% of the RV amplitude (the same numbers are $2.9$\,\ms\ 
and $4$\%, respectively, if we leave out the last data point). 
Furthermore, no correlation with the RV variation is seen. Therefore, 
we are confident that the source of the RV variation is stellar wobble 
due to the gravitational pull of a sub-stellar companion.

%
%
\section{Stellar and planetary parameters}
\label{sec:params}

The cornerstone of the derivation of the absolute parameters of a
planet discovered by radial velocity and transit observations is the
accurate estimation of the stellar mass and radius. 

The procedure often involves determination of \teff\ and [Fe/H], as
well as \logg, from high-resolution spectroscopic analysis. If an 
accurate distance is available, for example from a Hipparcos parallax, 
then the absolute magnitude can be used to improve the value for the 
stellar radius. In general, none of the methods yield better than 
$\sim 10$\% accuracy in the derived stellar parameters (implying 
corresponding limits to the accuracy of the absolute planet parameters). 
An improvement can be achieved by using the value of $a/R_\star$ 
derived from the transit light curve to measure directly the mean 
stellar density \citep{sozzetti07}. Since this latter parameter is 
directly linked to \logg, entering in the spectroscopic analysis, 
we followed an iterative determination of the stellar and planetary 
parameters.

%
%
\notetoeditor{This is the intended place of \tabr{stelpar}}

\begin{deluxetable}{lll}
\tabletypesize{\scriptsize}
\tablecaption{
	Stellar parameters for HAT-P-4.
\label{tab:stelpar}}
\tablewidth{0pt}
\tablehead{
	\colhead{Parameter} &
	\colhead{Value} &
	\colhead{Source}
}
\startdata
\teff\ (K)	    &	$5860\pm80              $  & SME\tablenotemark{a}\\
\feh\ (dex)	    &	$+0.24\pm0.08           $  & SME \\
\vsini\ (\kms)	    &	$5.5\pm0.5              $  & SME \\
Mass (\msun)	    &	$1.26\ [-0.14, 0.06] $  & Y$^2$+LC+SME\tablenotemark{b}\\
Radius (\rsun)	    &	$1.59\ [-0.07, 0.07] $  & Y$^2$+LC+SME\\
\logg\ (cgs)	    &	$4.14\ [-0.04, 0.01] $  & Y$^2$+LC+SME\\
$\lstar$ (\lsun)    &	$2.68\ [-0.34, 0.39] $  & Y$^2$+LC+SME\\
$M_{\rm{V}}$ (mag)  &   $3.74\ [-0.16, 0.16] $  & Y$^2$+LC+SME\\    
Age (Gyr)	    &	$4.2\  [-0.6,  2.6 ] $  & Y$^2$+LC+SME\\
Distance (pc)	    &	$310\pm30            $  & from M$_V$\tablenotemark{c}\\
\enddata
\tablenotetext{a}{SME = ``Spectroscopy Made Easy'' package to generate 
      synthetic spectra and to fit the observed ones \citep{valenti96}.}
\tablenotetext{b}{Y$^2$+LC+SME = Yonsei-Yale isochrones \citep{demarque04}, 
      transit light curve modeling and SME analysis.}
\tablenotetext{c}{Using $V=11.22\pm0.12$ of \cite{droege06} and assuming 
      zero reddening.} 
\end{deluxetable}

First, for the determination of \feh\ and \teff, we used the iodine-free
template spectrum from Keck. The modeling was performed using the SME
software \citep{valenti96} incorporating the same method and atomic
data as given in \cite{valenti05}. We obtained \teff$=6032\pm80$\,K, 
\feh$=+0.32\pm0.08$\,[dex] and \logg$=4.36\pm0.11$. 

Next, for the computation $a/\rstar$, we fitted the high precision 
KeplerCam light curve (see \figr{lcs}) by using the formulae of 
\cite{mandel02} with quadratic limb darkening coefficients from 
\cite{claret04}. We set $e=0.0$ as a result of our test on the RV data 
with non-zero eccentricity. Fitted parameters were the center of 
transit $T_{\rm c}$, the radius ratio $\rpl/\rstar$, the normalized 
relative semi-major axis $a/\rstar$, and the impact parameter $b$.

%
%
\notetoeditor{This is the intended place of \tabr{planet}}

\begin{deluxetable}{ll}
\tablecaption{
	\label{tab:planet}
	Spectroscopic, light curve and planet parameters of HAT-P-4}
\tablehead{
	\colhead{Parameter} &
	\colhead{Value}
} 
\startdata
Spectroscopic parameters:                  &  \\
Period (d)\tablenotemark{a}\dotfill        &    $3.056536\pm0.000057$\\
${\rm T}_{\rm c}$ (HJD)\tablenotemark{a}
\dotfill                                   &	$2,\!454,\!245.8154\pm0.0003$\\
                                           &	$2,\!454,\!248.8716\pm0.0006$\\
$K$ (\ms)\dotfill			   &	$81.1\pm1.9$\\
Offset velocity (\ms)\tablenotemark{b}\dotfill   &	$12.1\pm0.9$\\
$e$\tablenotemark{c}\dotfill		   &	$0.0$\\ \\
Light curve parameters:                    &  \\
Transit duration (day)\dotfill	 	   &	$0.1760\pm0.0003$\\
$a/R_\star$\dotfill                        &    $6.04\ [-0.18,0.03]$\\
$R_p/R_\star$ \dotfill                     &    $0.08200\pm0.00044$\\
$b\equiv a\cos i/R_\star$\dotfill          &    $0.01\ [-0.01,0.23]$\\ \\
Planet parameters:                         &  \\
\mpl (\mjup)\dotfill		           &    $0.68\pm0.04$ \\
\rpl (\rjup)\dotfill		           &	$1.27\pm0.05$ \\
\rhopl (\gcmc)\dotfill			   &	$0.41\pm0.06$ \\
\logg$_p$ (cm\ $s^{-2}$)\dotfill	   &    $3.02\pm0.02$ \\
$a$ (AU)\dotfill		           &	$0.0446\pm0.0012$ \\
\ipl (deg)\dotfill		           &    $89.9\degr\ [-2.2,0.1]$ \\	
\enddata
\tablenotetext{a}{Taken from the photometry (HATNet and 2 nights of \flwof\ ).}
\tablenotetext{b}{The $\gamma$ velocity is $-1.58\pm0.24$\,\kms, 
from the DS data.}
\tablenotetext{c}{Adopted}
\end{deluxetable}
%

Next, to compute the stellar mass and radius we relied on current 
stellar evolution models. As in our earlier papers, we compared the 
observational properties of the host star with a finely interpolated
grid of model isochrones from \cite{demarque04}. The inferred stellar 
properties are based on the best match to the measured values of 
$T_{\rm eff}$, \feh, and $a/\rstar$ within their observational 
errors, in a $\chi^2$ sense. This procedure led to a better approximation 
of the stellar gravitational acceleration with \logg$=4.12\pm0.04$. 

In the second loop of iteration we used the newly determined value 
of \logg\ in the SME analysis and redid the above sequence of computation. 
This led to a slightly modified set of stellar and planetary parameters 
with somewhat smaller errors than in the first loop. Since the changes 
were, in general, fairly small (e.g., \logg\ has changed to $4.14\pm0.03$), 
we decided to stop the iteration after this second loop.  
The final stellar and planetary parameters are shown in \tabr{stelpar} 
and \tabr{planet}, respectively. 

Concerning the derived parameters and their errors, we note the 
following. The dependence of the result on the evolutionary models 
was tested by using the isochrones given for solar-scaled $Z=0.03$ 
models with core overshooting by \cite{pietr06}. As noted by the 
authors, their models are hotter by some $200$~K than those of 
\cite{demarque04}. Therefore, we used an effective temperature 
of $6060$~K. With these input parameters we got nearly the same 
stellar and planet parameters as from the Yonsei-Yale isochrones 
(i.e., \rhopl~$=0.41$~\gcmc, age$=4.5$~Gyr). By using models 
without overshooting or of lower effective temperature, we got 
larger ages and also slightly larger densities, up to 
\rhopl~$=0.43$~\gcmc. The stability of the planet density is 
mainly related to the correlated change of stellar mass and radius 
when models or input parameters are changed. These age ranges and 
the derived metallicity fit reasonably to the relation recently 
given by \cite{reid07}. We also note that the derived radius is 
$34$\% larger than the one corresponding to an unevolved main-sequence 
star. This explains the longer than expected transit duration  
by which we were puzzled at the early phase of the discovery.

%
%
\section{Discussion and conclusions}
\label{sec:disc}

We presented the discovery data and derived the physical parameters 
of HAT-P-4b, an inflated planet orbiting BD+36~2593. Among the 20
transiting planets discovered so far, there are five with 
\rhopl~$\lesssim0.4$\,\gcmc. All others have at least $50$\% higher 
densities. For ease of comparison, \tabr{puffy} lists the relevant 
properties of the five inflated planets. It is remarkable how similar 
these planets are (except for TrES-4 that has distinctively low 
density). With its Safronov number of $0.036$, HAT-P-4b belongs to 
the Class II planets according to the recent classification of 
\cite{hansen07} and (together with TrES-4) further strengthens the 
mysterious dichotomy of the known transiting planets in this parameter. 
The parent star of HAT-P-4b is among the largest radii, largest mass, 
lowest gravity and highest metallicity transiting planet host stars. 

%
\notetoeditor{This is the intended place of \tabr{puffy}}

\begin{deluxetable}{lcccccc}
\tabletypesize{\scriptsize}
\tablecaption{
	Comparison of the properties of inflated
	planets.\tablenotemark{a}
\label{tab:puffy}}
\tablewidth{0pt}
\tablehead{
	\colhead{Name}          &
	\colhead{P}             &
	\colhead{a}             &
	\colhead{M}             &
	\colhead{R}             &
	\colhead{$\rho$}        &
	\colhead{$\log g$}     \\
	\colhead{}              &
	\colhead{(d)}           &
	\colhead{(AU)}          &
	\colhead{(M$_J$)}       &
	\colhead{(R$_J$)}       &
	\colhead{(\sc{cgs})}    &
 	\colhead{(\sc{cgs})} 
}
\startdata
WASP-1b    & 2.52 & 0.038 & 0.87 & 1.40 & 0.39 & 3.04 \\  
HAT-P-4b   & 3.06 & 0.045 & 0.68 & 1.27 & 0.41 & 3.02 \\
HD~209458b & 3.53 & 0.045 & 0.64 & 1.32 & 0.35 & 2.96 \\ 
TrES-4     & 3.55 & 0.049 & 0.84 & 1.67 & 0.22 & 2.87 \\
HAT-P-1b   & 4.47 & 0.055 & 0.53 & 1.20 & 0.38 & 2.96 \\ 
\enddata
\tablenotetext{a}
{Data from \cite{shporer07}, this paper, \cite{burro07}, \cite{mandu07}  
and \cite{winn07}. From top to bottom, metallicities for the parent stars 
are: $0.23$ \citep{stempels07}, $0.24$, $0.02$ $0.0$ (adopted) and $0.13$.}
\end{deluxetable}

Current models of irradiated giant planets are able to match the 
observed radii of most of the planets without invoking any additional 
heating mechanism. Higher metallicity cases, such as the present one, 
however, may pose problems (assuming that the planet and star have similar 
metallicities). More metals imply two opposite effects on the radius: 
(i) inflating it due to higher opacities in the envelope; 
(ii) shrinking it due to the higher molecular weight of the interior 
and the possible development of a large high density core. These 
effects have been discussed recently by \cite{burro07}. Since WASP-1b 
is similar in several aspects (i.e., irradiance, metallicity) to 
HAT-P-4b, we consider the coreless models of WASP-1b as shown in 
Fig.~7 of \cite{burro07}. It seems that HAT-P-4b can be fitted by 
near solar metallicity coreless models, assuming that its age is not 
too much greater that $4$\,Gyr. We also refer to the layered convective 
mechanism of \cite{chabrier07} that gives an alternative explanation 
for planets with inflated radii. 

We conclude that more definite statements on the relation of the 
observations and planet structure theories can be made only by reaching 
higher accuracy in the observed star/planet parameters. Nevertheless, 
HAT-P-4b (together with WASP-1b) does not seem to support the existence 
of a simple relation between host star metallicity and planet's core mass 
\citep[see][]{guillot06,burro07}.     

\acknowledgments
Operation of the HATNet project is funded in part by NASA grant
NNG04GN74G.
Work by G.~\'A.~B.\ was supported by NASA through Hubble
Fellowship Grant HST-HF-01170.01-A.
G.~K.~wishes to thank support from Hungarian Scientific Research
Foundation (OTKA) grant K-60750.
We acknowledge partial support from the Kepler Mission under NASA
Cooperative Agreement NCC2-1390 (D.~W.~L., PI).
%
G.~T.~acknowledges partial support from NASA Origins grant NNG04LG89G.
The Keck Observatory was made possible by the generous financial
support of the W.~M.~Keck Foundation. D.~A.~F is a Cottrell Science
Scholar of the Research Corporation. We acknowledge support from NASA 
grant NNG05G164G to DAF.

\end{document}